\documentclass[lettersize,journal]{IEEEtran}

\usepackage{amsmath,amsfonts}
\usepackage{algorithmic}
\usepackage{algorithm}
\usepackage{array}
\usepackage[caption=false,font=normalsize,labelfont=sf,textfont=sf]{subfig}
\usepackage{textcomp}
\usepackage{stfloats}
\usepackage{url}
\usepackage{verbatim}
\usepackage{graphicx}
\usepackage{cite}

\usepackage[utf8]{inputenc}
\usepackage{enumitem}
\usepackage{cprotect}
\usepackage{hyperref}
\usepackage{booktabs}
\usepackage[round-mode = places, round-pad = false,]{siunitx}
\usepackage{multirow}
\usepackage[normalem]{ulem}
\usepackage{tikz}
\usetikzlibrary{positioning, shapes.geometric, arrows, calc}

\newcommand{\HowManySignals}{five}
\newcommand{\NTotalMeas}{22}
\newcommand{\HWVersion}{1.0}
\newcommand{\NMeasPerCam}{3}
\newcommand{\NCams}{6}
\newcommand{\NFullCamRotations}{30}

\newcommand{\SubCaption}[1]{%
\parbox{0.75\textwidth}{\footnotesize
\vspace*{1em}

#1}}

\tikzstyle{input} = [trapezium, trapezium stretches, trapezium left angle=70, trapezium right angle=110, minimum width=2.5cm, minimum height=2em, text centered, draw=black, fill=blue!30]
\tikzstyle{output} = [trapezium, trapezium stretches, trapezium left angle=70, trapezium right angle=110, minimum width=2.5cm, minimum height=2em, text centered, draw=black, fill=green!30]
\tikzstyle{process} = [rectangle, minimum width=2.5cm, minimum height=2em, text centered, draw=black, fill=orange!30]
\tikzstyle{decision} = [diamond, minimum width=2.5cm, minimum height=2em, text centered, draw=black, fill=green!30]
\tikzstyle{arrow} = [thick,->,>=stealth]

\title{An Open Source Validation System for Continuous Arterial Blood Pressure Measuring Sensors}
\author{
    Attila R\'epai,
    S\'andor F\"oldi,
    P\'eter S\'otonyi,
    Gy\"orgy Cserey \IEEEmembership{Member, IEEE}
    \thanks{A.\ R\'epai is with Jedlik Innovation Ltd., Budapest, Hungary}
    \thanks{S. F\"oldi and Gy. Cserey are with Faculty of Information Technology and Bionics, P\'azm\'any P\'eter Catholic University, Budapest, Hungary}
    \thanks{P. S\'otonyi is with Faculty of Medicine, Department of Vascular and Endovascular Surgery, Semmelweis University, Budapest, Hungary}
}
\date{January 2023}

\begin{document}

\maketitle

\begin{abstract}

    Measuring the blood pressure waveform is becoming a more frequently studied area. The development of sensor technologies opens many new ways to be able to measure high-quality signals. The development of such an aim-specific sensor can be time-consuming, expensive, and difficult to test or validate with known and consistent waveforms. In this paper, we present an open source blood pressure waveform simulator with an open source Python validation package to reduce development costs for early-stage sensor development and research. The simulator mainly consists of 3D printed parts which technology has become a widely available and cheap solution. The core part of the simulator is a 3D printed cam that can be generated based on real blood pressure waveforms. The validation framework can create a detailed comparison between the signal waveform used to design the cam and the measured time series from the sensor being validated. The presented simulator proved to be robust and accurate in short- and long-term use, as it produced the signal waveform consistently and accurately. To validate this solution, a 3D force sensor was used, which was proven earlier to be able to measure high-quality blood pressure waveforms on the radial artery at the wrist. The results showed high similarity between the measured and the nominal waveforms, meaning that comparing the normalized signals, the RMSE value ranged from $0.0276 \pm 0.0047$ to $0.0212 \pm 0.0023$, and the Pearson correlation ranged from $0.9933 \pm 0.0027$ to $0.9978 \pm 0.0005$. Our validation framework is available at https://github.com/repat8/cam-bpw-sim. Our hardware framework, which allows reproduction of the presented solution, is available at https://github.com/repat8/cam-bpw-sim-hardware. The entire design is an open source project and was developed using free software.

\end{abstract}


\section{Introduction}

\IEEEPARstart{M}{easuring} the continuous noninvasive blood pressure (CNIBP) waveform is getting more important and more frequently studied. The main reasons behind it are the frequency of cardiovascular-related diseases in the population, the importance of ambulatory and patient monitoring and the development of sensor technologies. Development of CNIBP sensors requires a framework for reliable and reproducible evaluation. In early stages of development it is crucial to know the original signal waveform to be able to validate the signal measured by the developed sensor. But repeatable measurement cannot be completely guaranteed when working with human subjects, as their waveform can change significantly in short time periods that make the comparison evaluation a difficult task.
Also, it is difficult to find subjects with an abnormal blood pressure waveform that can lead to false conclusions when the sensor capabilities are evaluated.

For the above mentioned reasons, a CNIBP simulator would be a great tool to validate and evaluate the early stages of the development of a CNIBP sensor. Such devices were already presented in papers including solutions using pre-recorded or simulated signals to ones that simulate blood vessels as well. Most of these systems usually have significant costs.
A low cost, reliable simulator can solve this problem, and it can provide a useful tool for sensor development.

A frequent approach of blood pressure (BP) signal simulator solutions is to build an artificial wrist containing a model of the radial artery that is actuated pneumatically or hydraulically.
Validating MediWatch prototypes, Ng et al. \cite{Ng2004MediWatch} used a pneumatic pressure-pulse generator driven by a waveform simulation software that actuated a latex diaphragm placed on the artificial wrist's contact zone.
The contact zone of the artificial wrist can also be actuated with a linear motor, such as in the device of Heo et al. \cite{Heo2008SimLinear} designed for training oriental medicine practitioners and sensor validation.

Yang et al.\ developed a cam-based radial pulsation simulator \cite{Yang2019camfollower} that was used in \cite{Jun2020robotic} for the evaluation of a robotic pulse sensor. This simulator reproduces a single period of representative waveform, obtained via averaging 40 periods of waveform recording.
In their implementation a rotating cam is connected to a piston and the pressure wave is transduced pneumatically to an artificial radial artery embedded into a silicone wrist surface.
The authors measured 3.2\% phase delay and a repeatability of $CV = 0.23$\% and $CV = 0.82$\% for the heart rate and the pulse pressure, respectively.

McLellan et al.\ designed a BP pattern simulator \cite{McLellan2014threepoint} that is applicable for training practitioners of pulse diagnostics.
Therefore, it targets manual sensing with three fingers with a microprocessor-controlled solenoid for each.
The device also measures the gripping force with a pressure sensor placed at the opposite side, under the thumb.
They implemented a combination of three waveforms and four rhythms, but the article does not mention evaluation results.

Another approach is to closely reproduce the characteristics of human cardiovascular system.
The simulator Yang et al. \cite{Yang2019cardiovascular} consists of modules simulating the heart and the valves, the aorta with bifurcation (to generate reflected waves) and peripheral resistance including the wrist with the radial artery.
This complex system is capable generating age-dependent BP waveforms by simulating arterial stiffness as well.

Most of the above mentioned solution aims to be generally usable and simulate not just a signal waveform but also the measurement environment. Therefore, these solutions are complex and their implementation requires advanced manufacturing technologies.

In this manuscript, we introduce a simulator design that plays back a predefined signal using a mechanic transmission system and a validation framework in Python.
In the Methods section, we describe the implementation details and the method of evaluation. In the Results section we compare the ground truth signal to the one measurable on the device and in the Discussion section we propose some further improvements.

\section{Methods}

\subsection{Hardware design}

\begin{figure*}[!t]
    \centering
    \begin{tikzpicture}
        \node[anchor=south west,inner sep=0] (image) at (0,0) {\includegraphics[width=0.8\textwidth]{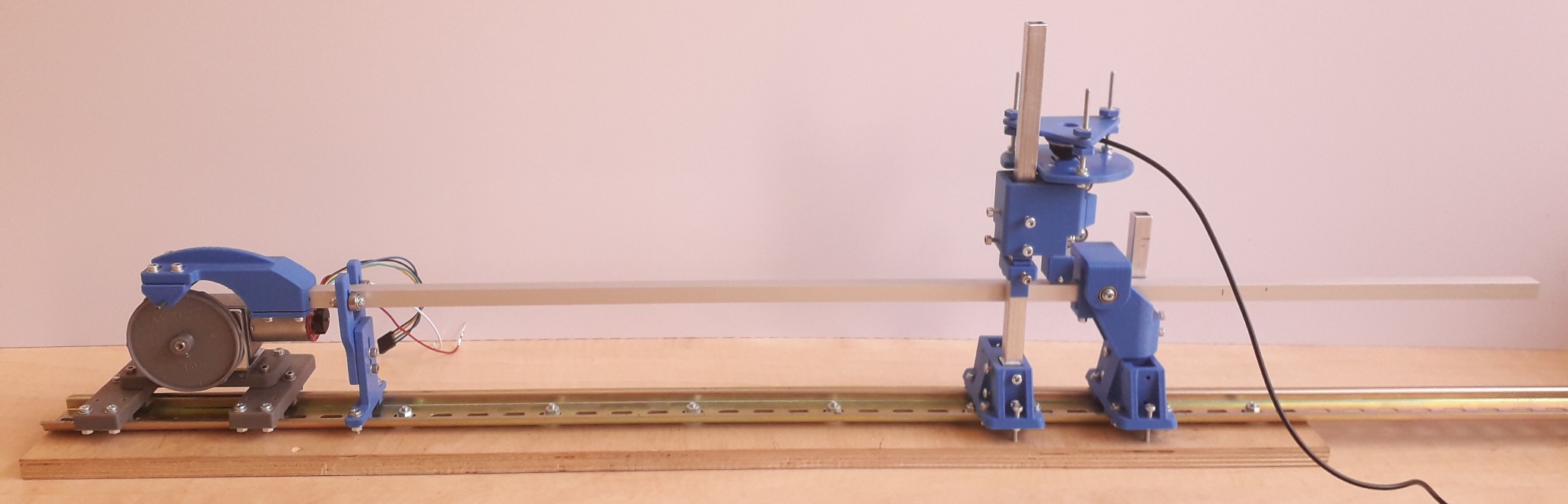}};
        \begin{scope}[x={(image.south east)},y={(image.north west)}]
            \node (O) at (0.03, 0.4) {$O$}; \draw[arrow] (O) -- (0.11, 0.31);
            \node (C) at (0.05, 0.9) {$C$}; \draw[arrow] (C) -- (0.09, 0.35);
            \node (J2) at (0.1, 0.9) {$J_2$}; \draw[arrow] (J2) -- (0.11, 0.41);
            \node (M) at (0.3, 0.9) {$M$}; \draw[arrow] (M) -- (0.18, 0.35);
            \node (L) at (0.5, 0.9) {$L$}; \draw[arrow] (L) -- (0.5, 0.42);
            \node (J1) at (0.95, 0.5) {$J_1$}; \draw[arrow] (J1) -- (0.71, 0.42);
            \node (J3) at (0.95, 0.7) {$J_3$}; \draw[arrow] (J3) -- (0.67, 0.44);
            \node (S) at (0.8, 0.9) {$S$}; \draw[arrow] (S) -- (0.68, 0.7);
            \node (T) at (0.6, 0.9) {$T$}; \draw[arrow] (T) -- (0.68, 0.55);
            \node (P) at (0.95, 0.9) {$P$}; \draw[arrow] (P) -- (0.68, 0.69);
            \node (Rh) at (0.4, 0.9) {$R_h$}; \draw[arrow] (Rh) -- (0.4, 0.17);
            \node (Rv) at (0.55, 0.9) {$R_v$}; \draw[arrow] (Rv) -- (0.64, 0.35);
        \end{scope}
    \end{tikzpicture}
    \caption{Parts of the simulator: an exchangeable cam $C$ with center $O$ rotated by a motor $M$; a lever $L$ between a revolute joint $J_1$ and the half joint of the cam follower $J_2$; a translating follower $T$ attached with $J_3$ at an adjustable horizontal position along $R_h$ in contact with sensor $S$ at point $P$; all parts mounted on the rail $R_h$.}
    \label{fig:parts-photo}
\end{figure*}
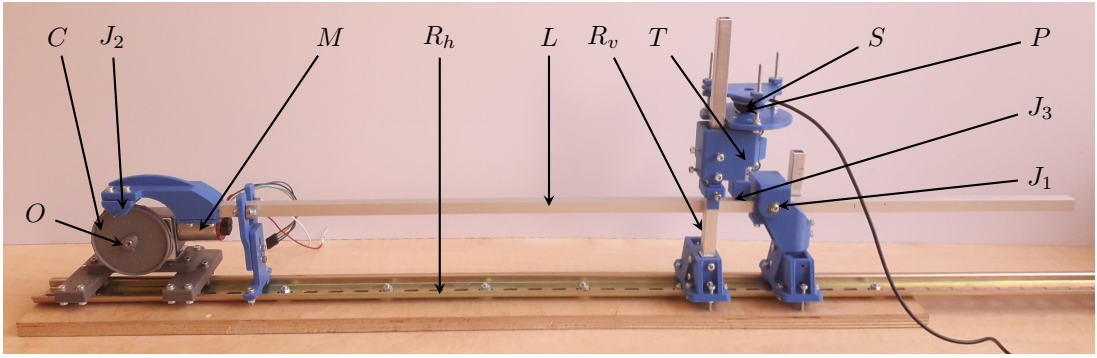

\noindent Cam-follower based designs are frequently used in a wide range of industrial applications, notably in valve control and mass production machines, as they are a reliable and easy to design solution for function generation \cite{NortonDesignOfMachinery}.
In the field of cardiovascular signal simulation, Yang \cite{Yang2019camfollower} used a similar approach, although they actuated a pneumatic system with the cam.
In the presented simulator design, shown in Fig.~\ref{fig:parts-photo}, the sensor to be tested is directly actuated by a planar linkage connected to the cam follower.

Keeping affordable production in mind, a modular design was implemented where all components are either 3D-printable or standard, widely available parts. The main elements of the simulator are 1) the waveform generator based on the cams, 2) a scaling mechanism in order to achieve a realistic (human-like) amplitude and 3) a sensor holder unit also responsible for a realistic pressure force level.
The design has to be able to emit waveforms with amplitudes of fraction of millimeters and must minimize the noise introduced by the system itself.
It also must be customizable in order to be suitable for custom sensors with different shape and size.

An exchangeable cam $C$, which provides the desired waveform, with a center $O$ connected to a motor $M$.
The waveform needs to be scaled to achieve realistic amplitude (details are in Section \ref{sec:protocol}) at the palpation point $P$.
This scaling is implemented with the lever $L$ leaning on the revolute joint (fulcrum) $J_1$ (equipped with rolling bearings to minimize the noise introduced by the joint) and the cam follower $J_2$.
The design of the fulcrum ensures that the top plain of $L$ is in level with the center of $J_1$, which simplifies the computation of the position of $J_3$ and thus $P$ given a camshaft angle $\theta$ (and the corresponding input level $h(\theta)$).
Depending on configuration, a class~1 or class~2 lever can be assembled from the same mechanical parts.
Class~1 lever is useful for cams with inverted signal.
This lever implements an oscillating cam follower.
The length $R_L$ of $L$ was chosen to be around 500~mm to allow convenient placement of the parts at a 20 times amplitude downscaling.
The physical limitation of scaling is due to the dimensions of the parts $J_1$ and $J_3$.
The motion of the lever is constrained to be in the vertical plane only, side movement noise is prevented by the fork $F$.

The cam followers were made of stainless steel pins fixed perpendicular to the cam plain, forming curved (mushroom) followers.
Cam followers are designed to be easy to customize; they are attached to the cam follower holder with screws.
The holder can support two cam follower heads: this design enables measuring the baseline rim on the cam (details are in Section \ref{sec:cam-design}).

The motion of the lever is transmitted to the sensor by a second, translating roller follower $T$ with vertical longitudinal axis.
This roller follower forms joint $J_3$.
This mechanism ensures that the sensor is only actuated along the $z$ axis.
The translating follower has an exchangeable contact head at the sensor at $P$, so that users can choose the optimal surface for their measurements.
The plate surrounding the contact head is also easy to replace, in order to allow additional components
like artificial skin or structures geometrically more similar to the human wrist.
The sensor's normal vector can be adjusted by three screws of the sensor holder, while the baseline pressure can be set by moving the module along its vertical rail $R_v$, setting the sensor holder's height $h_S$.
The basic sensor holder included in our assembly is a simple planar one, other sensors may need to be supported with a custom design, but the interface defined by the three elongated holes is easy to follow.

All parts are mounted on a DIN rail $R_h$, which allows easy assembly and modification and ensures proper alignment of the lever with respect to the cam.
The simulator has a modular design with 3D printed parts and standard metal profiles to facilitate reproduction of the device.

\subsection{Cam design}\label{sec:cam-design}

\begin{figure}[!t]
    \centering
    \includegraphics[width=\linewidth]{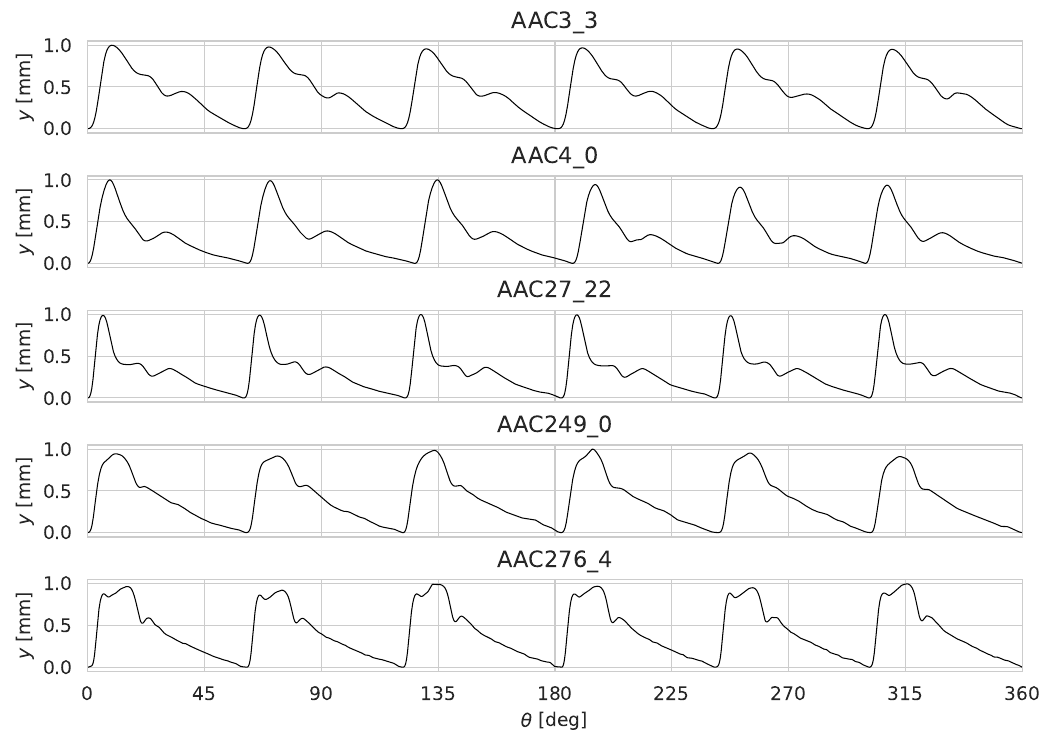}
    \cprotect\caption{Waveforms of the cams, as represented on the cam, amplitude vs.\ rotation angle.}
    \label{fig:cam-signals}
\end{figure}

\noindent The cam has an important role as it generates the blood pressure waveform. To construct it, noninvasive arterial blood pressure waveforms from the PhysioNet \cite{PhysioNet} "Autonomic Aging: A dataset to quantify changes of cardiovascular autonomic function during healthy aging" dataset \cite{2021AutonomicAging} (hereafter AAC) was used. From the 15 age groups presented in the dataset, we selected three noticeably different waveforms, all measured with the same device (CNAP 500, CNSystems; MP150, BIOPAC Systems) at $f_s = 1000$~Hz sampling frequency.
Table~\ref{tab:signal-selection} shows the details of the selected signals. AAC27 features 3-peak periods, with a clearly measurable reflected wave; AAC4 only has the systolic and dicrotic peaks; AAC3 has reflected waves without a preceding onset; AAC276 has reflected peaks preceding the systolic peaks; while AAC249 is a typical waveform of an aged individual (see Fig.~\ref{fig:cam-signals}).

\begin{table}
    \centering
    \caption{Signal Sample Selection from \cite{2021AutonomicAging}}
    \label{tab:signal-selection}

    \begin{tabular}{lcccc}
        \toprule
        Signal ID & Record ID & Age group & From $i$th onset \\
        \midrule
        AAC3 & 0003 & 45--49 & 3 \\
        AAC4 & 0004 & 30--34 & 0 \\
        AAC27 & 0027 & 18--19 & 22 \\
        AAC249 & 0249 & 80--84 & 0 \\
        AAC276 & 0276 & 75--79 & 4 \\
        \bottomrule
    \end{tabular}
\end{table}

As the original signals in some cases contained measurement noise or likely linearly interpolated sections, a Butterworth lowpass filter was applied with cutoff frequency $f_{max} = 30$~Hz.
Afterwards, a characteristic point detection algorithm was used introduced in Section \ref{sec:validation}.
Based on the onset points the baseline wander estimated by cubic spline interpolation was subtracted from the original signal.
This step also ensures a continuous signal even at the boundary of the selected section.
From the resulting filtered and baseline corrected signal $N = 6$ consecutive cardiac cycles were chosen.

This waveform was normalized between 0 and $y_{max}$ and added point-by-point to the radius of a circle,
so that the $N$ periods form a full cycle around the perimeter.
The radius chosen is $r = 30 \text{mm}$ at baseline.
Measurements were performed with $y_{max} \in \{0.75; 1; 1.5\} \text{mm}$.
Smaller amplitude has the advantage of smaller pressure angle $\phi$ and smaller lever length $r_1$, but may reach the accuracy limitations of 3D printing.

The resulting cam can be 3D printed.
A Python software package was developed that automates the aforementioned steps of building a cam based on an input waveform.
The software makes use of OpenSCAD code-based CAD tool \cite{OpenSCAD} to generate a 3D-printable STL file from the auto-generated SCAD code.

Two 3D printing technologies were tested: a CraftBot printer with PLA filament and Sonic Mini 8k printer with Phrozen resin (TR300).
CraftBot with PLA filament was used for printing the rest of the parts.

A baseline rim is added to the side of the cam.
It is a cylinder with a diameter similar to the baseline of the signal, its purpose is to evaluate cam quality with respect to printing distortions and positioning errors.
The cam follower has an adjustable head capable of reading the signal surface, the baseline rim or both (baseline with the peaks of the signal, useful for alignment).

The cams were rotated by a 12V DC motor (GW370 worm gear motor) at a nominal speed 10~rpm, approximating normal pulse rate.
The reason of our preference of a simple DC motor to a stepper motor is to reduce costs and complexity.
A DC power supply unit EB2025T (Thurlby-Thandar Instruments Ltd.) was used.
A 2~mm thick PDMS sheet was placed between the motor and the motor base in order to reduce high frequency vibration noise.
Similarly, PDMS or rubber dampers can be placed between the simulator board and the desk it is mounted on.

\subsection{Configuration}

\noindent Without developing new components, the user can control the output signal with the adjustable joints and the cam.
These parameteres are the cam profile; $r_1$, the distance of $J_1$ and $J_2$ along $L$ (the fulcrum can slide on $L$); $r_2$, the distance of $J_1$ and $J_3$ along $L$ ($R_v$ can slide on $R_h$); $\epsilon$ (eccentricity), the horizontal distance of $J_2$ and $O$ and $h_S$, the height of sensor along the vertical rail $R_v$. This way the baseline force can be adjusted.

The user can control supply voltage as well in the range 10~V--14~V to simulate an approximation of different heart rates with the same waveform.

\subsection{OptoForce OMD-20-SE-40N sensor}\label{sec:optoforce}

\noindent The OptoForce OMD-20-SE-40N is a 3D tactile force sensor\cite{2011TarCserey} of OnRobot (formerly OptoForce), consisting of a silicone rubber hemisphere with translucent fill and a reflective inner surface.
Deformation of this hemisphere is measured optically.
Infrared light is reflected onto multiple light sensing elements and the amount of light changes as the dome deforms.
From this, a 3D force vector is calculated and can be continuously viewed and logged on a computer.
It was proven that this sensor is suitable for human CNIBP waveform measurement following the principle of applanation tonometry~\cite{foldi2018novel}.

\subsection{Validation protocol}\label{sec:protocol}  

\noindent The measurement protocol defines a 3~min~20~s long recording using the OptoForce OMD-20-SE-40N sensor (see \ref{sec:optoforce}) the following way:

\begin{enumerate}[noitemsep]
    \item With the force sensor lifted, insert the signal cam.
    \item Set $f_s = 333$~Hz, lower the sensor and adjust tilt to minimize the $x$ and $y$ component of the measured force vector.
    \item Start the motor and adjust $h_S$ and $r_2$ so that realistic minimum sensor value (baseline amplitude) and signal amplitude, respectively, can be acquired. Based on prior measurement experience on human subjects, this means cca.\ 900--1100~units of sensor value and 100 -- 130~units of delta amplitude. We checked these values online via the OptoForce UI.
    \item Turn the cam to a position between the zero position (marked with a symbol \verb|<| on the printed model) and the dicrotic peak of the previous heart cycle. This allows simple comparison with the nominal signal; the first period of the measured signal can be detected with a simple threshold.
    \item Start the measurement and record 20~s from resting sensor. This section provides data for statistics of base noise level and wander of the sensor.
    \item Start the motor and record 3~min of simulated signal.
    \item Stop recording.
    \item Stop the motor.
\end{enumerate}

\subsection{Validation process}\label{sec:validation}

\noindent The validation process was performed with the help of our software package \verb|cam_bpw_sim|.

First, cam models were generated for the \HowManySignals{} signals introduced in \ref{sec:cam-design} and these were printed with 3D resin printer.
A mushroom cam follower of diameter $d_f = 1$~mm was used, simulator configuration was $r_1 = 500$~mm, $r_2 = 23$~mm, $\epsilon = 0$~mm, $U = 12$~V.
Then, sensor tilt and baseline pressure was set.
For each of the \HowManySignals{} cams, four measurement took place: one on the baseline rim of the cam with systolic peak marking for synchronization and the three measurements of the BP waveform.
Logged sensor data was processed offline.

\begin{figure}[!t]
    \centering
    \begin{tikzpicture}[node distance=1cm, every text node part/.style={align=center}]
        \footnotesize
        \node (csv) [input] {OptoForce CSV log};
        \node (fs) [input, right=of csv.east] {Predefined $f_s$};
        \node (full-signal) [process, below right=of csv] {Compute vector length, detect first onset};
        \draw [arrow] (csv) -- (full-signal);
        \draw [arrow] (fs) -- (full-signal);
        \node (raw-signal) [output, below of=full-signal] {Signal section};
        \draw [arrow] (full-signal) -- (raw-signal);
        \node (noise) [output, left=of raw-signal] {Noise section};
        \draw [arrow] (full-signal) -- (noise);
        \node (chpoints) [process, below of=raw-signal] {Detect characteristic points};
        \draw [arrow] (raw-signal) -- (chpoints);
        \node (nom) [input, below of=chpoints] {Nominal signal};
        \node (fcrs-long-bw) [process, below left=of nom] {Correct longterm BW; \\split to $N$ FCRs};
        \node (fcrs-bw) [process, below right=of nom] {Correct BW; \\split to $N$ FCRs};
        \draw [arrow] (chpoints) -| (fcrs-long-bw);
        \draw [arrow] (chpoints) -| (fcrs-bw);
        \draw [arrow] (nom) -- (fcrs-long-bw);
        \draw [arrow] (nom) -- (fcrs-bw);
    \end{tikzpicture}
    \caption{Flowchart of signal preprocessing.}
    \label{fig:preproc-flowchart}
\end{figure}
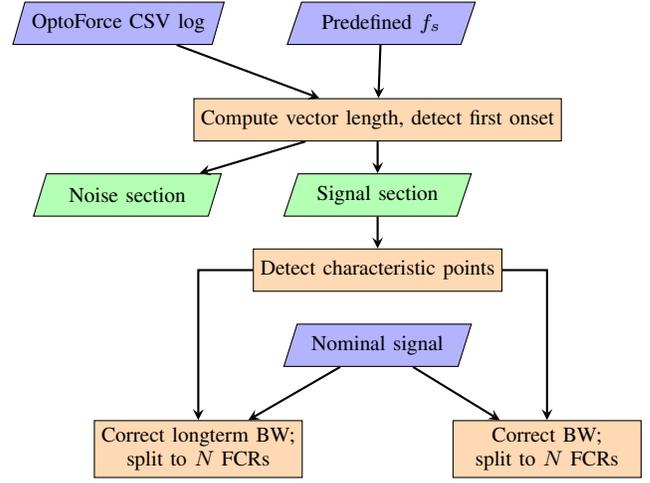

Prior to computing statistics, raw sensor data was preprocessed as shown in Fig.~\ref{fig:preproc-flowchart}.
Vector length was calculated from the $\vec{x}$, $\vec{y}$ and $\vec{z}$ force components of the log using the Euclidean norm and a time series was created with the nominal $f_s = 333$~Hz, assuming uniform sampling.
Following the protocol (introduced in Section \ref{sec:protocol}), the time series was split to noise and signal sections.
Then characteristic points were detected by a custom algorithm, in our software package \verb|ScipyFindPeaksDetector|:

\begin{enumerate}[noitemsep]
    \item Find rising edges with \verb|scipy.signal.find_peaks| on derivative.
    \item Find onsets as local minima. Between each onset:
    \begin{enumerate}[noitemsep]
        \item Find systolic peak as local maximum.
        \item Find reflected peak and dicrotic peak with \verb|find_peaks|.
        \item Find the onset of the reflected wave (if any) and the dicrotic notch as local minima.
    \end{enumerate}
\end{enumerate}

Statistics need two different transformed signals.

\begin{enumerate}
    \item A signal with the longterm baseline wander (BW) removed.
    This is accomplished by subtracting the cubic spline fit on the first onset of each full rotations of the cam (FCR).
    This type of baseline wander can be attributed to the sensor, removing it makes FCRs comparable but preserves cam baseline error for quality assessment.
    \item A signal with full baseline correction by subtracting the cubic spline fit on all onsets.
    This transformation cancels the baseline error of the cam, attributed to printing distortion (non-uniform scaling and inner tensions), making the waveform comparable to the nominal one.
\end{enumerate}

The nominal signal was transformed to match the sampling frequency and the amplitude of the measured signal.
Each of these signals were split to FCRs using an algorithm based on cross-correlation with the matched nominal signal.
Cam rotations were further split to cardiac cycles based on the onsets of the nominal signal.
In both levels, a sufficiently large margin was added to the signal sections to ensure that the first and the last onset point is included, even with the uncertainty of the point detection.
These margins were not involved in waveform comparison, only in statistics on characteristic points.

This validation system can evaluate the accuracy, the precision, the sensitivity, the robustness and the durability and reliability of a sensor, or in the case of this manuscript, these properties of the simulator itself, comparing it to a validated sensor.

Accuracy of the simulator was evaluated with root mean squared error (
\(\text{RMSE}(y, y_0) = \sqrt{\frac{1}{N} \sum_{i=1}^N (y_i - y_{0_i})^2}\), where $N$ is the number of samples, $y_i$ is a single measured value, $y_0$ is the corresponding nominal value) and Pearson correlation on pairs of measured cam rotations and the nominal signal and on pairs of measured cardiac cycles and the corresponding nominal cardiac cycle waveforms.
Additionally, the standard deviation of the characteristic points was computed.
RMSE and STD provides values in the original scale of the measurement, therefore these could be compared to the noise observed on the unactuated sensor.

Pearson correlation coefficient was calculated using \verb|scipy.stats.pearsonr|.
Precision of the system was evaluated by calculating RMSE on each pair of raw cam rotations, as well as on pairs of corresponding characteristic points.
Durability was assessed with a one hour long measurement using AAC27 where error with respect to time was calculated.

In total \NTotalMeas{} measurements were performed using simulator hardware version \HWVersion{},
consisting of \NMeasPerCam{} consecutive measurements and a base rim measurement with each of the \NCams{} cams (see \ref{tab:cam_summary},
and additional \NMeasPerCam{} measurements with AAC27 prior to the durability measurement.
Each measurement contained cca.\ \NFullCamRotations{} full cam rotations.

\begin{table*}
    \centering
    \caption{Cam and Measurement Properties}
    \label{tab:cam_summary}

    \begin{tabular}{llS[round-precision=1]S[round-precision=1]lS[round-precision=0, table-format=4]S[round-precision=0, table-format=4]}
    \toprule
    {Cam} & {Signal} & {$y_{max}$} & {$r_f$} & {Material} & {Meas.\ ampl.} & {\# meas.}\\
    \midrule
    AAC249\_Phr2 & AAC249\_0 & 1.0 & 0.0 & Phrozen & 117.5948126500529 & 3\\
    AAC276\_Phr2 & AAC276\_4 & 1.0 & 0.0 & Phrozen & 119.28238584219362 & 3\\
    AAC27\_Phr1 & AAC27\_22 & 1.0 & 0.0 & Phrozen & 117.14467025492434 & 3\\
    AAC27\_rf50\_Phr1 & AAC27\_22 & 1.0 & 0.5 & Phrozen & 113.83417532631587 & 3\\
    AAC3\_Phr1 & AAC3\_3 & 1.0 & 0.0 & Phrozen & 119.84290540676453 & 3\\
    AAC4\_Phr1 & AAC4\_0 & 1.0 & 0.0 & Phrozen & 119.12995753408252 & 3\\
    \bottomrule
    \end{tabular}
    \SubCaption{$y_{max}$ [mm] is the amplitude of the pitch curve on the cam, $r_f$ [mm] is the radius of the cam follower the cam is designed for, meas.\ ampl.\ [sensor unit] is the mean pulse pressure amplitude of the measurements and \# meas.\ is the number of measurements with the same cam.}
\end{table*}

Reliability of the system was evaluated on an approximately one hour long measurement with cam AAC27.
Longterm baseline wander was estimated with a cubic spline fit on the first onset of each FCR.
After baseline correction a short section at both ends were discarded as they may be impacted with boundary effects.
Then within a rolling window of $N = 30$ FCRs cross-RMSE and the RMSE against the nominal signal was computed.

Additional \NMeasPerCam{} measurements took place to evaluate the effect of different supply voltages.

\subsection{Software}

\noindent Two software packages were developed in the Python programming language.
First, package \verb|bpwave|\cite{bpwaveGithub} provides a generic signal representation data structure, \verb|bpwave.Signal|, that stores the time series, characteristic points, marks (named indices) and measurement metadata.
This package is published on the Python Package Index~\cite{bpwave} under MIT license and meant for general purpose usage in projects processing ABP signals.

Second, package \verb|cam_bpw_sim| \cite{camBpwSimGithub} contains the algorithms specifically designed to support simulator development, such as generating cams, evaluating measurements and scripts for reproducing the results presented in this article.
The package has a low level and a high level API, the latter also accessible via a command line application.


The application supports generating cams from arbitrary time series in the \verb|cam_bpw_sim.cam| module.
Measured or constructed signals can be preprocessed and transformed according to the desired amplitude and cam follower parameters.
Given the pitch curve coordinates $K: (x, y)$, calculated as adding signal amplitude $S(\theta)$ to cam baseline radius $r_C$, the coordinates of the cam profile are computed as
\begin{equation}
    \left\{ \begin{aligned}
        x_P &= x - r \cdot \frac{\text{d}y/\text{d}\theta}{\sqrt{(\text{d}x/\text{d}\theta)^2 + (\text{d}y/\text{d}\theta)^2}} \\
        y_P &= y + r \cdot \frac{\text{d}x/\text{d}\theta}{\sqrt{(\text{d}x/\text{d}\theta)^2 + (\text{d}y/\text{d}\theta)^2}}
    \end{aligned} \right. ,
\end{equation}
where $\theta$ is the rotation angle \cite{YiMechanisms}.

Quality check is implemented to warn the user if the signal is not suitable for a cam with the provided parameters (undercutting happens).
Besides the printable 3D model of the cam, the software saves all parameters and inputs for later usage in evaluation.


The \verb|cam_bpw_sim.signal| module provides extensions to the aforementioned \verb|bpwave| package.
These include normalization, resampling, denoising, characteristic point detection (see \ref{sec:validation}) and correction of baseline wander.
Interfaces are provided to allow users implement their own characteristic point detection and baseline correction algorithms as drop-in replacement of our implementations in the evaluation code.


The results presented in this article were calculated using the \verb|cam_bpw_sim.val| module.
This module contains algorithms to process measurement files such as splitting continuous measurements performed on the simulator to full cam rotations and cardiac cycles and measuring accuracy and precision.
The validation workflow is implemented in Jupyter notebooks, in order to allow easy customization and inspection of intermediary results.


The application saves cams, signals and measurement files in a well-defined structure together with metadata needed for reproducing validation results.
These are stored in HDF5 and JSON file formats for easier interoperability with other software products like MathWorks' MATLAB.

\section{Results}

\noindent Unit of RMSE in the following sections is sensor output values as measured by the OptoForce sensor, unless otherwise specified.


First, accuracy of the simulated waveforms was determined in terms of RMSE and Pearson's correlation coefficient $\rho$, comparing entire waveforms of baseline corrected full rotations of the cam to the nominal waveform.
These statistics are summarized in Table~\ref{tab:accuracy-wf}.
$RMSE_{rel}$ ranged from $0.0276\pm 0.0047$ to $0.0212 \pm 0.0023$ and $\rho_{rel}$ from $0.9933 \pm 0.0027$ to $0.9978 \pm 0.0005$.

\begin{table*}
    \centering
    \caption{Accuracy of the Simulated Waveform Compared to the Nominal Signal}
    \label{tab:accuracy-wf}

    \begin{tabular}{lS[round-precision=0, table-format=4]S[round-precision=4]S[round-precision=4]S[round-precision=4]S[round-precision=4]S[round-precision=4]S[round-precision=4]}
    \toprule
    {Cam} & {\#Comp} & {$\overline{{E}}$} & {$\sigma(E)$} & {$\text{{med}}(E)$} & {$\overline{{E_{{rel}} }}$} & {$\sigma(E_{{rel}})$} & {$\text{{med}}(E_{{rel}})$}\\
    \midrule
    AAC249\_Phr2 & 96.0 & 2.493683459310016 & 0.2687136289477065 & 2.432718541570839 & 0.021206517885664026 & 0.0022922453832636886 & 0.020733297854803684\\
    AAC276\_Phr2 & 93.0 & 2.984760720667732 & 0.4181053834254527 & 2.870067991211433 & 0.02502526719043909 & 0.0035391086370699827 & 0.024036260682495047\\
    AAC27\_Phr1 & 96.0 & 3.2371110290491587 & 0.5464515453961161 & 3.099409287571648 & 0.027631146282654857 & 0.00465600093819429 & 0.02651945298156656\\
    AAC27\_rf50\_Phr1 & 96.0 & 2.7657479851868594 & 0.26417352690566254 & 2.7259266649406992 & 0.024294358805740605 & 0.0023358963077806955 & 0.023865055548867965\\
    AAC3\_Phr1 & 94.0 & 3.1782398660468236 & 1.7183963506817275 & 2.107773856246141 & 0.026516609007573584 & 0.01432925090969605 & 0.01758573673549249\\
    AAC4\_Phr1 & 97.0 & 2.6607465225266007 & 0.22885075890465403 & 2.6117955359617175 & 0.02233477863024915 & 0.0019215059265557937 & 0.02196437748111656\\
    \bottomrule
    \end{tabular}
    \SubCaption{\#Comp is the total number of comparisons, $E$ is the error as defined by $\text{RMSE}(\text{measured FCR}, \text{nominal waveform})$, in original units and $E_{rel}$ is the relative error compared to the mean pulse pressure amplitude.}
\end{table*}

Second, characteristic points of the measured and the nominal signal were compared, with respect to time and amplitude.
Note that this evaluation may also be affected by the accuracy of the point detection algorithm,
as characteristic points may be ambiguous in the case of flat peaks and shoulders.

\begin{figure*}[!t]
    \centering
    \includegraphics[width=\textwidth]{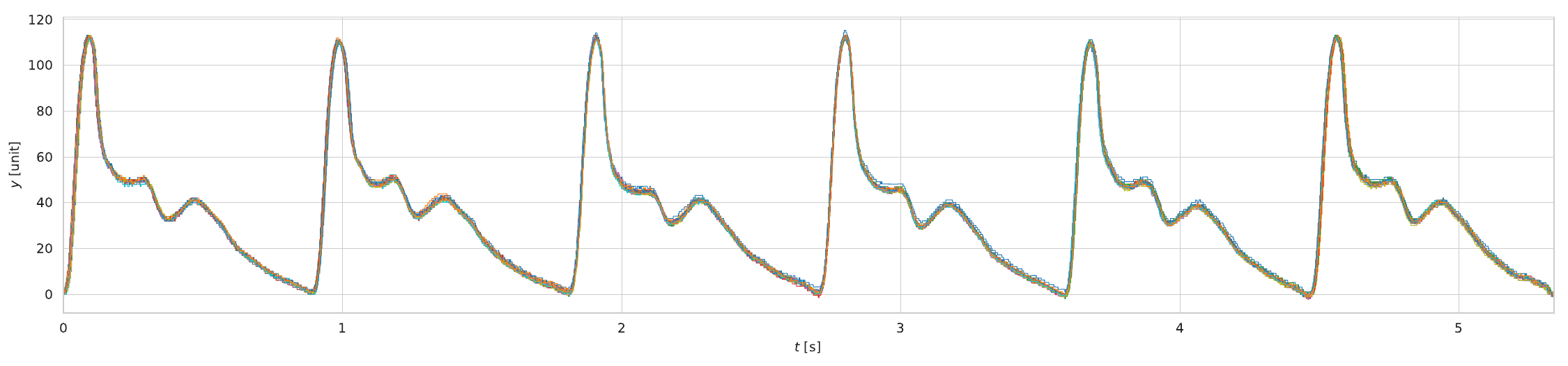}
    \cprotect\caption{32 full rotations of the cam AAC27 aligned with cross-correlation after longterm BW removal. Different colors mean the consecutive cycles.}
    \label{fig:fcrs-no-long-bw}
\end{figure*}

Precision was measured by the cross-comparison of full cam rotations without the longtime baseline wander (Fig.~\ref{fig:fcrs-no-long-bw}).
RMSE was between $0.0127 \pm 0.0036$ and $0.0182\pm 0.0077$,
Pearson's $\rho$ between $0.996557 \pm 0.0035$ and $0.9989 \pm 0.0008$.
These are summarized in Table~\ref{tab:precision}.

\begin{table*}
    \centering
    \caption{Precision Measurements: Cross Comparison of Full Cam Rotations}
    \label{tab:precision}

    \begin{tabular}{lS[round-precision=0, table-format=4]S[round-precision=4]S[round-precision=4]S[round-precision=4]S[round-precision=4]S[round-precision=4]S[round-precision=4]}
    \toprule
    {Cam} & {\#Comp} & {$\overline{{E}}$} & {$\sigma(E)$} & {$\text{{med}}(E)$} & {$\overline{{E_{{rel}} }}$} & {$\sigma(E_{{rel}})$} & {$\text{{med}}(E_{{rel}})$}\\
    \midrule
    AAC249\_Phr2 & 1488.0 & 1.7111132167048473 & 0.5841849400079865 & 1.565378561729299 & 0.014552509546065684 & 0.004974206727722765 & 0.013306902841059804\\
    AAC276\_Phr2 & 1396.0 & 1.773753806941379 & 0.639820007537236 & 1.6240460030057924 & 0.014872070167733446 & 0.005378991378350835 & 0.01363290867600175\\
    AAC27\_Phr1 & 1489.0 & 2.1271574647360385 & 0.9026860833787674 & 1.8940635593917217 & 0.01815467513016346 & 0.007698799867650716 & 0.016207456541078507\\
    AAC27\_rf50\_Phr1 & 1489.0 & 1.8686865298623805 & 0.6697939178923857 & 1.7416419253977675 & 0.016413144096560996 & 0.00589682074026634 & 0.015297474784170314\\
    AAC3\_Phr1 & 1427.0 & 1.5773460858698247 & 0.5367661979819679 & 1.4474662653090675 & 0.013161427720407742 & 0.004476837862704647 & 0.012073614961139296\\
    AAC4\_Phr1 & 1520.0 & 1.5101174589380413 & 0.43124088658528803 & 1.416286539181108 & 0.01267611483810576 & 0.0036207832123561537 & 0.011895212159561207\\
    \bottomrule
    \end{tabular}
    \SubCaption{\#Comp is the total number of comparisons (without comparison to itself), $E$ is the error (RMSE),  in original units and $E_{rel}$ is the relative error compared to the mean pulse pressure amplitude.}
\end{table*}

Regarding sensitivity, the capabilities of the 3D printers were determined using the baseline rim of the printed cams.
The diameter of this cylinder was measured at the onsets and for resin cams it was $58.02 \pm 0.03$~mm, for PLA cams $57.88 \pm 0.06$~mm compared to the nominal 58~mm.
Concave waveform parts in this setup must have a radius of curvature greater than 0.5~mm.
In the case of rising edges approximating vertical, an inverted cam design is recommended with class 1 lever setup.

\begin{figure*}[!t]
    \centering
    \includegraphics[width=\textwidth]{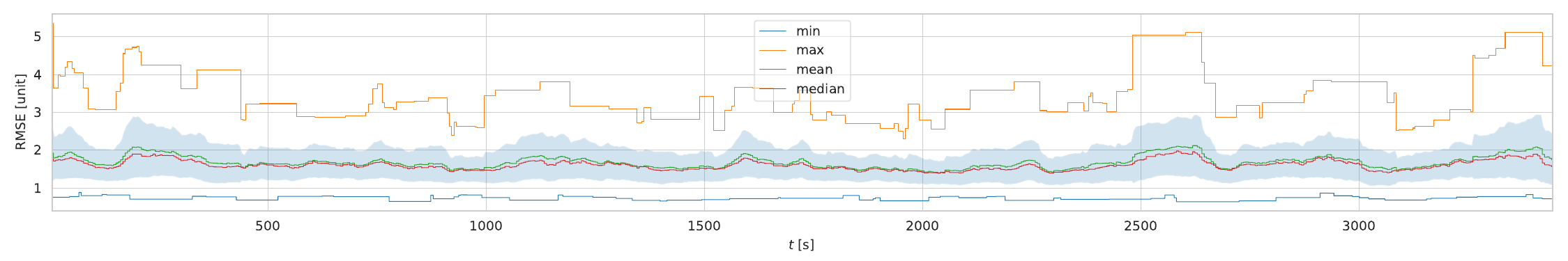}
    \caption{Rolling cross-RMSE of the longterm measurement wrt.\ time.}
    \label{fig:durability-rcrmse}
\end{figure*}

As for reliability, median cross-RMSE of the full measurement was $\approx 1.6$ units of amplitude.
The median of the cross-RMSE of the rolling window of 30 FCRs varied between 1.5 and 2 and fits a constant level over time (see Fig.~\ref{fig:durability-rcrmse}, confirming that precision of the system does not significantly change in time.
As for accuracy, rolling RMSE against the nominal signal held a median level $\approx 3.5$ units then after 2500~s, error started to increase.
This can be attributed to the slowly decreasing amplitude over time due a sensor-specific issue and the decreasing motor speed due to the power supply.

The proposed system allows simulating BP waveforms with different speeds, controlled by the input voltage of the DC motor.
Measurements on the selected AAC27 signal showed that there is no significant difference in precision when voltage is changed in the range 10--14~V, but accuracy is the best at the nominal 12~V (see Table~\ref{tab:acc-voltage}).

\begin{table*}
    \centering
    \caption{Effect of Input Voltage on Accuracy}
    \label{tab:acc-voltage}

    \begin{tabular}{rS[round-precision=0, table-format=4]S[round-precision=4]S[round-precision=4]S[round-precision=4]S[round-precision=4]S[round-precision=4]S[round-precision=4]}
    \toprule
    {$U$} & {\#Comp} & {$\overline{E}$} & {$\sigma(E)$} & {$\text{med}(E)$} & {$\overline{E_{rel}}$} & {$\sigma(E_{rel})$} & {$\text{med}(E_{rel})$}\\
    \midrule
    10.0 & 75.0 & 3.1973077224165625 & 0.45138431290940956 & 3.2091140336843633 & 0.026972515314096366 & 0.003807469362511575 & 0.027058106391071347\\
    11.0 & 88.0 & 3.055622278333701 & 0.5457322718405062 & 2.879265146516138 & 0.025884402863547165 & 0.004628580319519623 & 0.024398654987386505\\
    12.0 & 97.0 & 2.9988723414129352 & 0.4427823596126442 & 2.86599394218154 & 0.02547157421183533 & 0.0037584472079260386 & 0.024359920946675627\\
    13.0 & 107.0 & 3.144278684244267 & 0.38493313511860383 & 3.0755601648651476 & 0.026856343494068992 & 0.0032936793197212294 & 0.026291126846804604\\
    14.0 & 112.0 & 3.2501864195420316 & 0.4545200757620576 & 3.1846096130868213 & 0.02777144150778422 & 0.003882507842446474 & 0.02723528449535525\\
    \bottomrule
    \end{tabular}
    \SubCaption{\#Comp is the total number of comparisons (without comparison to itself), $E$ is the error (RMSE),  in original units and $E_{rel}$ is the relative error compared to the mean pulse pressure amplitude.}
\end{table*}

\section{Discussion}

\noindent The simulator presented here is designed to aid the early stages of development of continuous noninvasive arterial blood pressure sensors based on applanation tonometry.
The simulator outputs a vertical movement following a waveform determined by the cam.
The mechanical parts in direct contact with the sensor under evaluation are designed to be easy to customise,
in particular the pressure tip can be modified to have different contact surface, providing a single-point or linear contact, such as planar, hemispherical or cylindrical, the latter resembling the protruding surface of the pulsating surface of the radial artery.
The pressure tip can be coated in elastic material such as PDMS as well,
to emulate the damping effects of tissues between the arterial wall and the sensor.
The choice of surface depends on the shape of the sensor and on the conditions to be tested.
It is also important that simulated measurement produce sensor values that fall into the range that can be measured on human subjects.
Therefore, since different sensors may have different sensitivity, the simulator allows easy configuration of baseline amplitude and delta amplitude by the adjustment of $h_S$ and $r_2$, respectively.
The cam surface is highly customizable as well:
besides representing baseline-corrected BP waveforms,
software package \verb|cam_bpw_sim| also supports generating cams from arbitrary signals that may even contain intentionally introduced baseline wander.
In this latter case, the baseline wander correction step must be skipped at the evaluation process.
The number of cardiac cycles within a full rotation of the cam depends on two factors:
the rotation speed of the motor and the feasibility conditions of fabricating the cam, namely the pressure angle and curvature criteria.

The simulator was validated using the force sensor OMD-20-SE-40N because it is already proven in \cite{foldi2018novel} that it is suitable for tonometric measurement of continuous blood pressure waveform on the radial artery with high precision.
However, known issues of the sensor affect the measurement results presented here,
namely measurement noise and the slow decrease of measured signal amplitude (long-term baseline wander).
Consequently, the error metrics presented in this article incorporate the printing error as well as the measurement noise and baseline wander of the OptoForce sensor.

The results of the evaluation measurements suggest that the continuous blood pressure waveform simulator is capable of reproducing different types of pre-recorded waveforms with sufficient precision and accuracy for both longer (45~min) and shorter (2--5~min) timeframes covering the monitoring and diagnostic requirements.
Average noise amplitude measurable with the force sensor on a still surface is 3--6~units and the sensor outputs integer values.
Therefore, the precision and accuracy error presented is comparable to the sensor noise.

A major limitation of the presented simulator design is that the quality of the simulated waveform is highly dependent of the quality and accuracy of 3D printing of the cams.
Printing from PLA filament allows durable cams with smooth surface, but the belt-driven printer head results in distortion of the printed part.
However, this effect can be approximated and compensated with a non-uniform scaling, obtained from printing and measuring a specimen object before printing the cam.
Such specimen object is included in \cite{camBpwSimHwGithub}.
Stereolithography (photopolymer resin printing), on the other hand, is accurate in shape, but the surfaces perpendicular to the XY plane, such as the cam surface are less smooth and subject to pixel rounding errors and other printing and cleaning artifacts.

In the case of both tested technologies, accuracy was dependent on the complexity and feature-richness of the desired waveform.
It was observed that using resin printing, the waveform sections between the dicrotic peak and the next onset is affected by printing noise to the higher extent.
Besides printability, the curvature of the circular waveform must be less than the radius of the cam follower at all points, that is, if the waveform to be represented does not conform to this condition, then the solution can be to include less cardiac cycles and rotate the cam at a higher speed.
Our software provides functionality to help the user find problematic parts on the cams that may be corrected with printing with different settings or manual postprocessing.

The quality of the simulated waveform, notably accuracy over time, could be improved by using a stepper motor, however that increases production costs.

\section{Conclusion}

\noindent A hardware and software were successfully implemented to provide a validation environment to aid the development of continuous noninvasive arterial blood pressure sensors.
The simulator is easily configurable to adapt sensors with different measurement range and precision,
enabling test measurements with amplitude similar to that of human subjects.
The software package lets developers manufacture cams for various waveforms using rapid prototyping technology.
Following the Open Science initiative, the entire design is an open-source project and was developed using free software.

The simulator was validated using \NCams{} cams representing human ABP waveforms of different age groups and features.
Measurements performed with the OptoForce OMD-20-SE-40N sensor showed that the desired waveforms can be reproduced with an error comparable to the measurement error of the sensor.

\section*{Acknowledgments}
\noindent This work has been partially supported by the Ministry of Culture and Innovation of Hungary from the National Research, Development and Innovation Fund, financed under the 2020-1.1.5-GYORSÍTÓSÁV funding scheme (project no. 2020-1.1.5-GYORSÍTÓSÁV-2021-00022). \\ This work has been partially supported by the Ministry of Culture and Innovation of Hungary from the National Research, Development and Innovation Fund, financed under the TKP2021-NKTA funding scheme (project no. TKP2021-NKTA-66).

\bibliographystyle{IEEEtran}
\bibliography{bibliography}

\begin{IEEEbiographynophoto}{Attila Répai}
received the M.S.\ degree in computer science engineering from the Faculty of Information Technology and Bionics at Pázmány Péter Catholic University, Budapest, Hungary, in 2020. He is currently a researcher at Jedlik Innovation Ltd., Budapest, Hungary. His research interests include blood pressure monitoring and pulse diagnosis.
\end{IEEEbiographynophoto}

\begin{IEEEbiographynophoto}{Sándor Földi}
received the Ph.D.\ degree in information science from the Faculty of Information Technology and Bionics at Pázmány Péter Catholic University, Budapest, Hungary, in 2020. He is currently a researcher at the Faculty of Information Technology and Bionics, Pázmány Péter Catholic University, Budapest, Hungary. His research interests include blood pressure monitoring, pulse diagnosis, biomedical signal processing, and biomechanics.
\end{IEEEbiographynophoto}

\begin{IEEEbiographynophoto}{Péter Sótonyi}
MD received the Ph.D.\ degree in medical science from the Doctoral College at Semmelweis University, Budapest, Hungary, in 2004.  He specialized in surgery (2001), vascular surgery (2004) and health insurance (2010). He is currently the professor and head of department at the Faculty of Medicine, Department of Vascular and Endovascular Surgery, Semmelweis University, Budapest, Hungary. His research interests include blood pressure monitoring, etiopathology of aortic aneurysms, carotid interventions and cerebrovascular circulation, biomechanics of homografts.
\end{IEEEbiographynophoto}

\begin{IEEEbiographynophoto}{György Cserey}
received the Ph.D.\ degree in Infobionics from Faculty of Information Technology and Bionics at Pázmány Péter Catholic University (PPCU), Budapest, Hungary, in 2006. He is currently a professor at PPCU. In 2003 he was a visitor researcher at Notre Dame University (Indiana, USA). His research interests include sensory robotics, parallel computing, machine learning and neuro-bio-inspired systems.
\end{IEEEbiographynophoto}

\end{document}